\pdfoutput=1
\documentclass[twocolumn]{aastex63}
\submitjournal{ApJL}

\usepackage{graphicx}
\usepackage{txfonts}
\usepackage{hyperref}
\usepackage{textcomp}
\usepackage{gensymb}

\shorttitle{Subsurface zonal and meridional flows}
\shortauthors{Getling, Kosovichev, and Zhao}

\begin{document}

\newcommand{\vdag}{(v)^\dagger}
\newcommand\aastex{AAS\TeX}
\newcommand\latex{La\TeX}
\newcommand{\rem}{\textcolor{red}}

\title{Evolution of Subsurface Zonal and Meridional Flows in Solar Cycle 24 from Helioseismological Data}

\correspondingauthor{A. V. Getling}
\email{A.Getling@mail.ru}

\author{Alexander V. Getling}
\affil{Skobeltsyn Institute of Nuclear Physics, Lomonosov Moscow State
	University, Moscow, 119991 Russia}

\author{Alexander G. Kosovichev}
\affiliation{New Jersey Institute of Technology, NJ 07102, USA}

\author{Junwei Zhao}
\affiliation{W.W. Hansen Experimental Physics Lab., Stanford University, Stanford, CA 94304, USA}

\begin{abstract}
The results of determinations of the azimuthal and meridional velocities by time--distance helioseismology from Helioseismic and Magnetic Imager (HMI) onboard Solar Dynamics Observatory (SDO) from 2010 May to 2020 September at latitudes from $-60\degree$ to $+60\degree$ and depths to about 19 Mm below the photosphere are used to analyze spatiotemporal variations of the solar differential rotation and meridional circulation. The pattern of torsional oscillations, or latitudinal belts of alternating `fast' and `slow' zonal flows migrating from high latitudes toward the equator, is found to extend in the time--latitude diagrams over the whole time interval. The oscillation period is comparable with a doubled solar-activity-cycle and can be described as an extended solar cycle. The zonal-velocity variations are related to the solar-activity level, the local-velocity increases corresponding to the sunspot-number increases and being localized at latitudes where the strongest magnetic fields are recorded. The dramatic growth of the zonal velocities in 2018 appears to be a precursor of the beginning of activity Cycle 25. The strong symmetrization of the zonal-velocity field by 2020 can be considered another precursor. The general pattern of poleward meridional flows is modulated by latitudinal variations that are similar to the extended-solar-cycle behavior of the zonal flows. During the activity maximum, these variations are superposed with a higher harmonic corresponding to meridional flows converging to the spot-formation latitudes. Our results indicate that variations of both the zonal and meridional flows exhibit the extended-solar-cycle behavior, which is an intrinsic feature of the solar dynamo.

\end{abstract}

\keywords{Sun: differential rotation --- Sun: meridional circulation --- Sun: helioseismology}

\section{Introduction} \label{sec:intro}

The discovery of solar differential rotation based on sunspot observations was likely made by Scheiner as early as 1630, and the first determination of the rotation law were done more than two centuries later by \citet{Carrington_1863}. It was substantially improved, in particular, by \citet{Newton_Nunn_1951}, who also used sunspots as tracers of solar rotation, and by \citet{Snodgrass_1983}, who tracked solar magnetic fields. Further progress in these studies was related to Doppler measurements \citep{Howard_etal_1983, Snodgrass_Ulrich_1990}, tracking magnetic features \citep[e.g.,][]{Meunier_1999,Zhao_etal_2004}, and applying helioseismological inversions \citep[e.g.,][]{Thompson_etal_1996}.

Currently, differential rotation is considered a substantial component of nearly all solar-dynamo mechanisms, which greatly strengthens investigators' interest in an adequate knowledge and understanding of this phenomenon. Modern theoretical models of the solar differential rotation are based on the idea of interaction between the general rotational field and rotation-affected anisotropic turbulence, originally put forward by \citet{Lebedinsky_1941} \citep[see, e.g., a review by][]{Kitchatinov_2005}.

Based on Doppler observations, \citet{Howard_Labonte_1980} discovered a pattern of traveling torsional waves with alternating latitudinal zones of slow and fast rotation. These zones were found to originate near the poles and drift to the equator, where they disappear, with a period of about 22 years (which later came to be known as the \emph{extended solar cycle}). The entire pattern is, to a first approximation, symmetric about the equator. Further studies of the torsional waves were done by \citet{Labonte_Howard_1982}, \citet{Snodgrass_1991} (who analyzed the rotation of magnetic fields), etc. Helioseismological data revealed the pattern of torsional oscillations to extend throughout a wide depth range in the convection zone \citep{Kosovichev_1997,Howe_etal_2000,Vorontsov_etal_2002,Kosovichev_Pipin_2019}.

As shown by \citet{Lebedinsky_1941} and \citet{Kippenhahn_1963}, the differential rotation inevitably entails a meridional circulation. In essence, these are two closely related phenomena; in other words, two manifestations of a unique process. Such a flow with a velocity of order 10~m\,s$^{-1}$, directed poleward in each hemisphere, was detected by Doppler measurements \citep[][etc.]{Duvall_1979,Hathaway_etal_1996}, by tracing small magnetic features \citep{Komm_etal_1993} and sunspots \citep[e.g.,][who reported, however, smaller velocity values]{Howard_Gilman_1986}, and by helioseismological techniques \citep[][etc.]{Giles_etal_1997,Basu_etal_1999}.

Using the local helioseismic technique of ring-diagram analysis applied
to Michelson Doppler Imager (MDI) data from the Solar and Heliospheric Observatory \citep{Scherrer_1995}, \citet{Haber_2002} discovered that the meridional flow within the upper convection zone can develop additional circulation cells whose boundaries wander in latitude and depth as the solar
cycle progresses.  \cite{Zhao_Kosovichev_2004}, analyzing time--distance helioseismological measurements and inversions, detected torsional oscillations and meridional flows with a velocity of order 20~m\,s$^{-1}$ directed poleward in both hemispheres during the whole period of observations (1996--2002) and superposed with a higher harmonic corresponding to flows converging toward the activity belts. \citet{GonzalezHernandez_2010} found that, like the torsional oscillations, the extra meridional circulation cells started to develop at mid-latitudes three years prior to Solar Cycle 24. Further local helioseismology analysis of the solar-cycle variations of zonal and meridional flows in the upper convection zone was performed by \citet{Zhao_etal_2014,Kosovichev_2016,Komm_2014,Komm_2018,Lin_2018}. Using the ring-diagram inversion results, \citet{Komm_etal_2020} studied the solar-cycle variation of subsurface flows for two subsets that represent active and quiet regions for the period of 1996--2017 and argued that the bands of fast and slow zonal and meridional flow associated with quiet regions may represent precursors of the surface manifestation of magnetic activity of a solar cycle.

The presence of meridional circulation also constitutes an important ingredient of dynamo theories \citep{Charbonneau_2020}. Specifically, the  flux-transport dynamos include the meridional circulation as the basic mechanism of flux transport. \citet{Pipin_Kosovichev_2019} constructed a nonlinear mean-field dynamo model, with magnetic-field effects on angular-momentum and heat transport, which explained the extended 22-year cycle of torsional oscillations. The model also predicts the extended-cycle behavior of zonal variations of the meridional circulation.

Here, we analyze the patterns of zonal velocity (in essence, differential rotation) and meridional circulation at various levels below the photosphere and their evolution during the whole Solar Cycle 24, 2010--2020, using the subsurface-flow maps derived by time--distance helioseismology..

\section{The Data and the Processing Techniques Used}\label{obs}

The data of helioseismological determinations of the azimuthal and meridional velocity components, $v_x$ and $v_y$, that we use here are available from the Joint Science Operations Center (JSOC) of the Solar Dynamics Observatory \citep{Pesnell_2012}. The subsurface flow maps for the whole visible surface of the Sun are routinely produced every 8 hours by the time--distance helioseismology pipeline \citep{Zhao_Couvidat_2012} 
from the Helioseismic and Magnetic Imager (HMI) Dopplergrams \citep{Scherrer_2012,Schou_2012}. The flow maps used in our analysis are calculated for a grid of $1026\times 1026$ points spanning over 123\degree\ of heliographic latitude, $\varphi$, and longitude, $\lambda$, with a spatial resolution of 0\degree.12 and a time cadence of 8~hours. The flow maps are produced for the following eight characteristic levels below the photosphere, $d$ (the corresponding depth ranges for which the inversions were done are parenthesized): (0) $d=0.50$ (0--1)~Mm, (1) $d=2.00$ (1--3)~Mm, (2) $d=4.00$ (3--5)~Mm, (3) $d=6.00$ (5--7)~Mm, (4) $d=8.50$ (7--10)~Mm, (5) $d=11.50$ (10--13)~Mm, (6) $d=15.0$ (13--17)~Mm, (7) $d=19.0$ (17--21)~Mm.

We consider a period from 2010 May to 2020 September. The procedure of our analysis of the helioseismological data is as follows. First, for a given time and a given depth level, we average the $v_x$ and the $v_y$ field over latitude within narrow latitudinal zones, or belts, centered at 41 latitude, from $-60\degree$ to $+60\degree$ with a step of 3\degree. The chosen belt widths are 7\degree.68, $3\degree.84$, or $1\degree.92$. Thus, each full-disk $x$- or $y$-velocity field is represented by a set of $1026\times 41$ mean values. Second, we average the obtained mean belt values over longitude and then calculate their running averages $U_x$ and $U_y$ for the entire data set with a time window of 45, 183, or 365 days and a 10-day step. Third, from each resultant field, we subtract the field averaged over the whole period, i.e., the mean zonal velocity $\langle U_x\rangle$ in the case of $U_x$ and the mean meridional velocity $\langle U_x\rangle$ in the case of $U_y$, to obtain deviations from the mean, $(U_x)_\mathrm d$ and $(U_y)_\mathrm d$. We compare the flow variations with the monthly sunspot number and the monthly averaged vertical (radial) magnetic fields, $B_z$.

\begin{figure*} 
	\centering
\includegraphics[width=\textwidth]{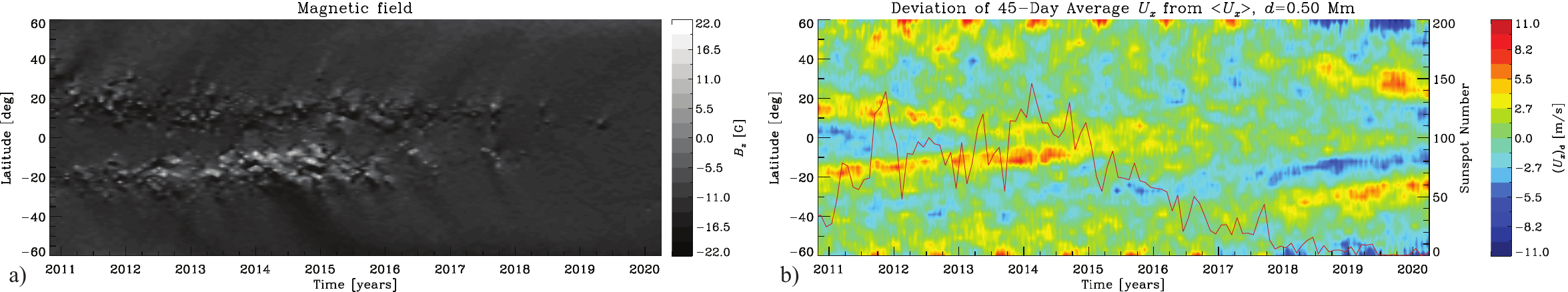}
	\caption{Time--latitude diagrams: (a) monthly averaged $B_z$; (b) the deviation, $(U_x)_\mathrm d$, of the 45-day running zonal-velocity average, $U_x$, from its mean over the whole interval considered, $\langle U_x\rangle$, at $d=0.50$~Mm. The red curve shows the variation of the monthly sunspot number.
		\label{fig1}}
\end{figure*}

\begin{figure*} 
	\centering
\includegraphics[width=\textwidth]{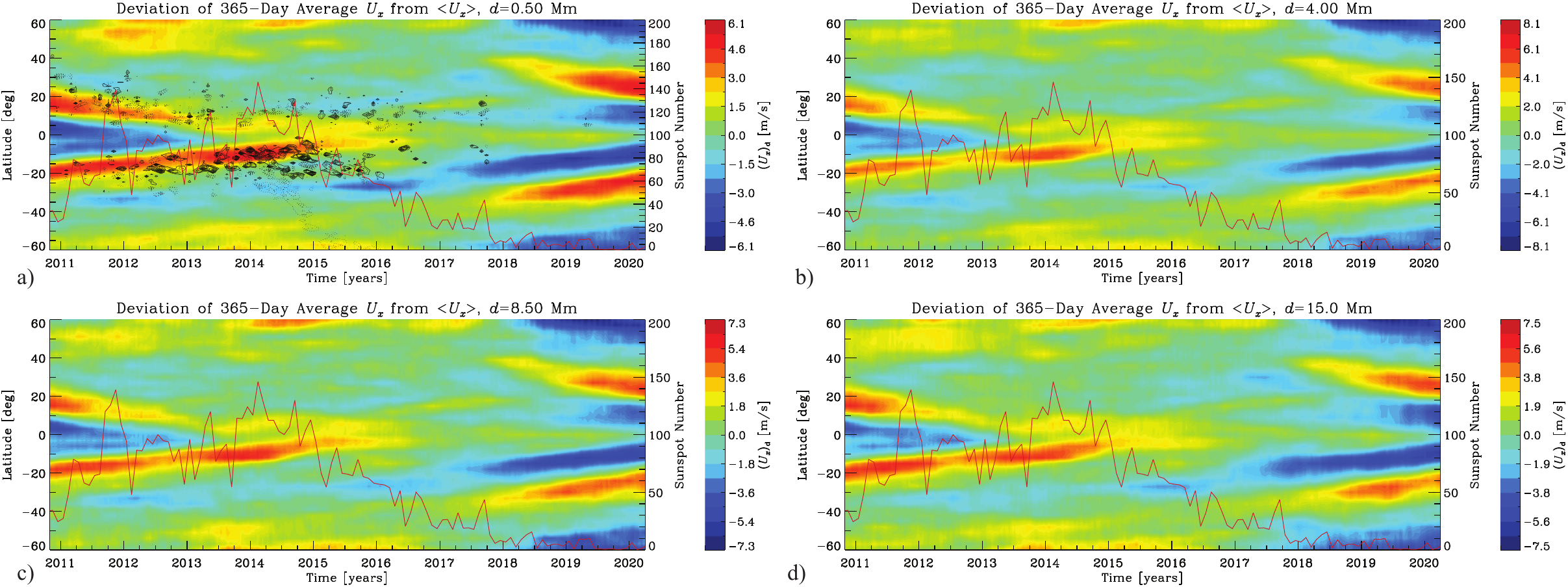}
	\caption{Time--latitude diagrams of $(U_x)_\mathrm d$ computed for the 365-day running average, $U_x$, at levels $d=0.5$, 4.0, 8.5, 15.0~Mm. Contours of monthly-averaged $B_z$ (solid for positive and dotted for negative values) are superposed in panel (a). The red curve in each diagram shows the variation of the monthly sunspot number.
		\label{fig2}}
\end{figure*}

\section{Results}\label{results}

Varying the belt width in the above-mentioned range was found to have only minor effects on the averaged zonal and meridional flows. For this reason, we present here only the results obtained with a belt width of 1\degree.92, or about 23.3~Mm, which basically smooths the supergranulation pattern but does not substantially smear the latitudinal dependencies. As for the running-averaging time window, we give below only one diagram obtained with a 45-day-long window and argue that one-year averaging is of most interest.

\begin{figure*} 
	\centering
\includegraphics[width=\textwidth]{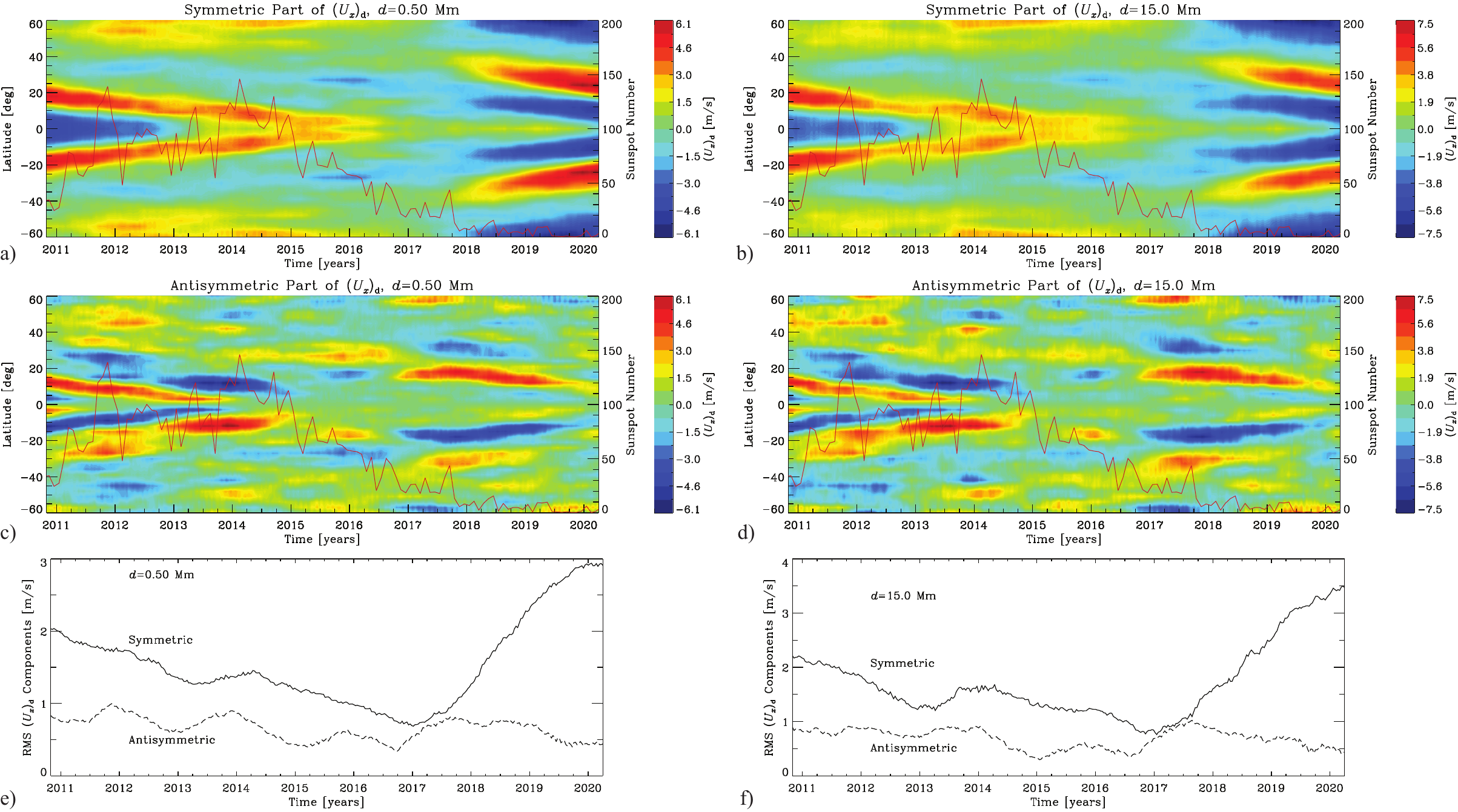}
	\caption{Time--latitude diagrams of the $(U_x)_\mathrm d$ components: symmetric (a--b), and antisymmetric (c--d) with respect to the equator, $(U_x)_\mathrm d^\mathrm s$ and $(U_x)_\mathrm d^\mathrm a$, at levels $d=0.5$ and 15.0~Mm. The red curve in each diagram shows the variation of the monthly sunspot number. Time variations in the rms $(U_x)_\mathrm d$ symmetric (e) and antisymmetric (f) components at levels $d=0.5$ and 15.0~Mm.
		\label{fig3}}
\end{figure*}

\subsection{Differential rotation}

The spatiotemporal variations in the zonal velocity are illustrated in time--latitude diagrams representing the deviation of the running-averaged velocity from its mean over the entire period considered, $(U_x)_\mathrm d=U_x-\langle U_x\rangle$ (Figures~\ref{fig1}b and ~\ref{fig2}). The variations of solar activity are indicated by the red curve of sunspot number, by a time--latitude diagram of $B_z$ in Figure~\ref{fig1}a, and by contours of $B_z$ in Figure~\ref{fig2}a (solid contours show positive values and dotted contours show negative values).

Figure~\ref{fig1}b shows $(U_x)_\mathrm d$ for the 45-day running average. As can be seen from this diagram, the pattern at high latitudes is modulated with a pronounced one-year period. The velocity oscillations in the two hemispheres are in antiphase. This modulation effect is attributable to variations in the inclination angle of the Sun's rotational axis. Since these oscillations are not a physical effect stemming from the internal dynamics of the convection zone, we eliminate them using a 365-day window to compute running averages $U_x$. From here on, we consider only 365-day running averaging applied to $U_x$ (and $U_y$).

The corresponding deviations, $(U_x)_\mathrm d$, of $U_x$ from the mean are shown in the colored maps of Figure~\ref{fig2} for levels $d=0.5$, 4.0, 8.5, and 15.0~Mm.
The patterns of spatiotemporal variations in $(U_x)_\mathrm d$ are very similar at all depths from the photosphere to about 15~Mm. In all diagrams, long wedge-like features are present, which consist of alternating bands of increased and decreased velocities. The bands of like-sign $(U_x)_\mathrm d$ located nearly symmetrically about the equator migrate from relatively high latitudes to the equator. One pair of the high-speed bands stretches over the whole 10-year time interval, from  about $\pm 20\degree$ latitude in 2010, at the left edge of the map, to the equator at the beginning of 2020, near its right edge. The velocities in this pair reach a maximum in the southern band at the epoch of a solar-activity maximum, not far from the beginning of 2014. Not only does this main maximum coincide in time with the sunspot-number peak, but it also agrees well in its latitudinal location with the maximum of magnetic fields. In the diagram for $d = 15.0$~Mm, another local maximum in the southern band is also distinguishable in the mid-2011, during a secondary activity peak.

The bands of `fast' rotation are interleaved with bands of reduced speed, or `slow' rotation. The latter are also located almost symmetrically about the equator and also converge to the equator. A second pair of the increased-velocity bands converging to the equator can be traced since 2011--2012 in the southern hemisphere and, with much less certainty, since 2018 in the northern hemisphere. The bands of reduced velocities located in between the bands of increased velocities stretch throughout the entire period considered. On the whole, the pattern of velocity variations in 2010--2020, as revealed at all the depths, reflects the presence of global zonal flows (torsional oscillations) in the outer envelope of the solar convection zone. This pattern is consistent with the previously detected 22-year periodicity of solar torsional oscillations, or the extended solar cycle.

\begin{figure*} 
	\centering
\includegraphics[width=\textwidth]{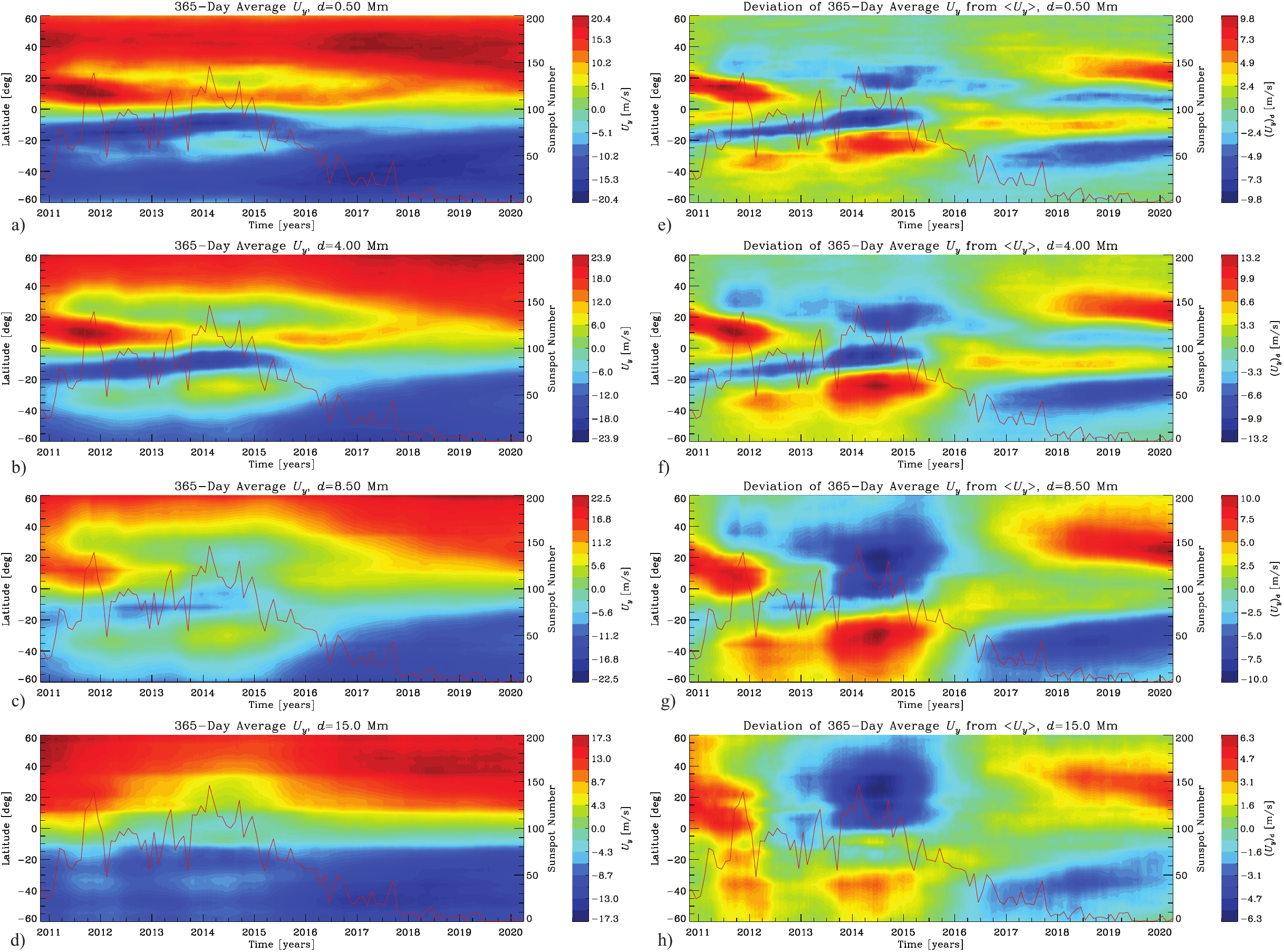}
	\caption{Left: time--latitude diagrams representing the 365-day running average of the meridional velocity, $U_y$, at levels~$d=0.5$, 4.0, 8.5, 15.0~Mm; right: the deviation, $(U_x)_\mathrm d$, of $U_y$ from its mean over the whole interval considered, $\langle U_y\rangle$, at the same levels.
		\label{fig4}}
\end{figure*}

There is no strict symmetry with respect to the equator in the distribution of velocities, the latter being typically larger in the southern hemisphere (Figure~\ref{fig2}). To quantify this asymmetry, which varies widely, we represent the $(U_x)_\mathrm d$ field for each time as the sum of the components $(U_x)_\mathrm d^\mathrm s$ and $(U_x)_\mathrm d^\mathrm a$ symmetric (even) and antisymmetric (odd) with respect to the equator, calculate their RMS values (which characterize the amplitudes of the two components), and treat these values as functions of time.

The time variation of these quantities is displayed in Figure~\ref{fig3} for levels $d=0.5$ and 15.0~Mm. In the mid-2017, shortly before the years of activity minimum, 2018--2019, the symmetric and antisymmetric components have comparable, relatively small rms values (Figures~\ref{fig3}e--f). During the minimum, in 2018, a dramatic growth of the zonal velocities in the second pair of fast-rotation bands and in the pair of lower-latitude slow-rotation bands is observed. This acceleration appears to be a precursor of the beginning of Cycle 25. The strong symmetrization of differential rotation by the beginning of the year 2020 can be considered another precursor. If we keep in mind that the symmetric-to-antisymmetric amplitude ratio exhibited a systematic decline in the years 2010--2017 (totally, by a factor of about 2), we can conjecture that maximum values of this ratio are typical of the early solar-cycle stages and minimum ones, of the years preceding the activity minima.

\subsection{Meridional circulation}

We deal with the data for meridional flows in a quite similar manner, being interested in both the running-averaged velocity $U_y$ and its deviation from the mean, $(U_y)_\mathrm d=U_y-\langle U_y \rangle$. Time--latitude diagrams of these fields for four levels are presented in Figure~\ref{fig4}. To identify details of the flow pattern described below, we compare the $U_y$ and $(U_y)_\mathrm d$ fields shown in the left and the right column of Figure~\ref{fig4}.

\begin{figure*} 
	\centering
\includegraphics[width=\textwidth]{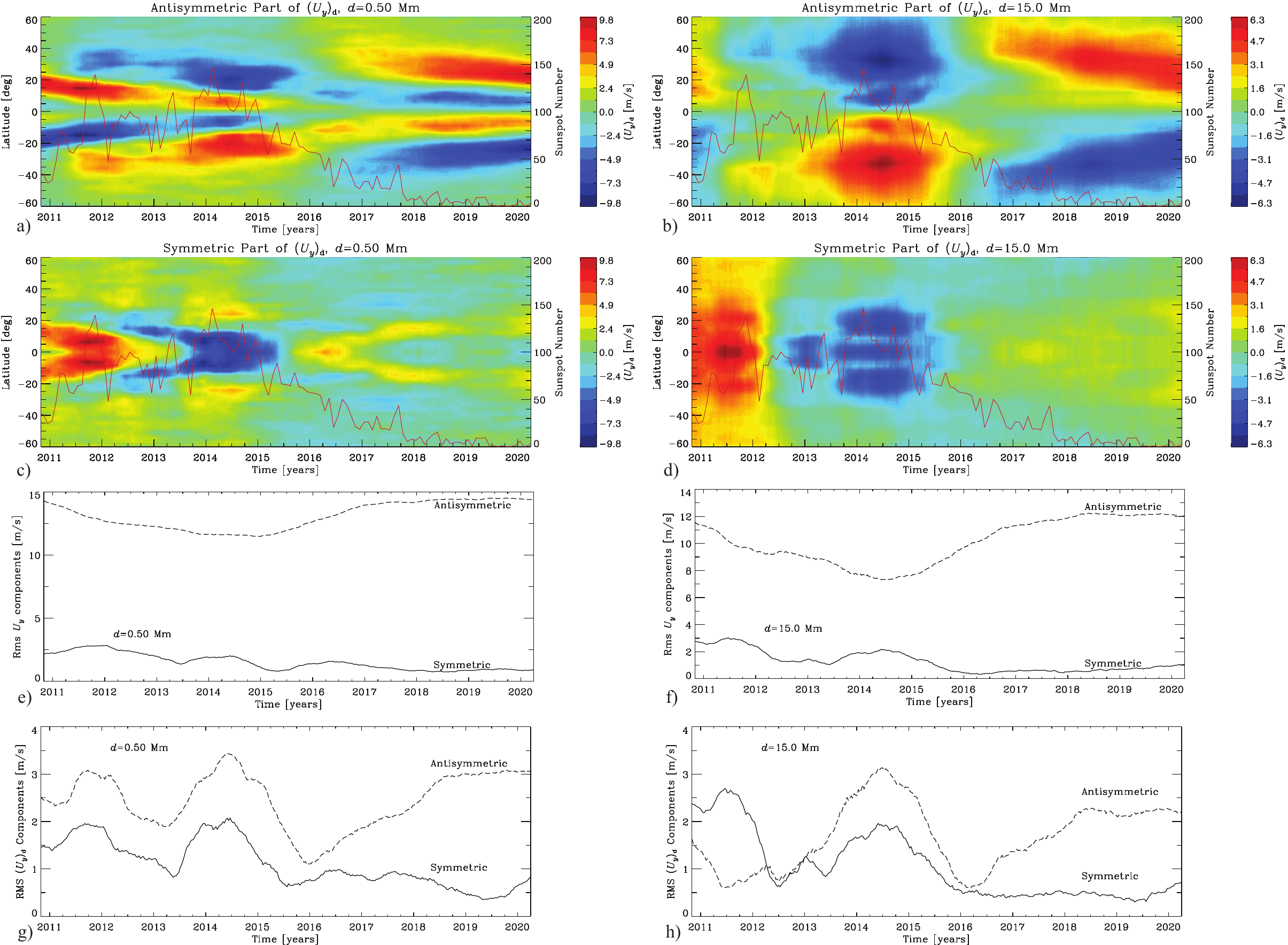}
	\caption{Time--latitude diagrams of the $(U_y)_\mathrm d$ components: antisymmetric (a--b) and symmetric (c--d) with respect to the equator at levels $d=0.5$ and 15.0~Mm computed for the 365-day running average, $U_x$. The red curve in each diagram shows the variation of the monthly sunspot number. Time variations in the rms $U_y$ (e--f) and rms $(U_y)_\mathrm d$ (g--h) of the symmetric (solid curves) and antisymmetric (dashed curves) components at levels $d=0.5$ and 15.0~Mm.
		\label{fig5}}
\end{figure*}

In these maps, a particular feature of interest is present in 2014, near the activity maximum. Specifically, against the background of the flow directed poleward in both hemispheres, there are reduced-velocity areas centered near $\pm 20\degree$ and corresponding to the highest-activity period (cf. them with the areas of strong magnetic fields in Figures~\ref{fig1}a and \ref{fig2}a). At level $d=0.5$~Mm, the velocity does not change its sign in these areas, i.e., the decelerated meridional flows are directed poleward, as outside the particularity. However, this is not the case at levels $d=4.0$ and 8.5~Mm, where reverse flows are observed. Near a latitude of $-10\degree$, the meridional flow at levels $d=0.5$--8.5~Mm is, on the contrary, accelerated. A similar increased-velocity area in the northern hemisphere can be noted only at level $d=0.5$~Mm as a small yellow island in the $(U_y)_\mathrm d$ map (Figure~\ref{fig4}a). Thus, during the year of activity maximum, the general pattern of poleward meridional flow is superposed with a higher harmonic corresponding to meridional flows directed toward the highest-activity latitudes. This flow component is more pronounced in the southern hemisphere; at levels $d=4.0$ and 8.5~Mm (Figures~\ref{fig4}b, c), it manifested itself in a reverse, equatorward flow at relatively high latitudes that encountered a regularly directed flow at lower latitudes.

On the whole, the above-described feature can be traced over a longer time interval. At level $d=0.5$~Mm, signs of reduced speeds in the northern hemisphere are notable starting from the secondary sunspot-number peak in the mid-2011, until the mid-2017 (the stretched yellow area in the $U_y$ map mainly occupying latitudes near $\varphi \sim 20\degree$ and a wider latitudinal range during the highest-activity interval, see Figure~\ref{fig4}a). In 2011, a temporary increase in the poleward-flow velocity was observed at levels $d=0.5$ and 4.0~Mm and latitudes of 5--20\degree\ (Figures~\ref{fig4}a, b). In the southern hemisphere, a similar area of accelerated poleward flow at the same levels is less pronounced and stretches from the beginning of 2011 to mid-2013 (from $-20\degree$ to $-10\degree$ latitude), where it joins to the above-mentioned well-defined area of accelerated flow in 2014. At level $d=8.5$~Mm, the areas of reverse flows extend over wider latitudinal ranges than at $d=4.0$~Mm.

Thus, the depth variation of the meridional-flow pattern indicates that a secondary flow converging to the spot-formation latitudes was present in the subphotospheric layers at depths of $d=4.0-8.5$~Mm during the period of the highest activity.

The time variation of the $(U_y)_\mathrm d$ components symmetric and antisymmetric about the equator is illustrated in Figure~\ref{fig5} along with the behavior of the symmetric and antisymmetric components of the rms $U_y$ and rms $(U_y)_\mathrm d$. The $U_y$ field is basically antisymmetric (corresponding to the poleward flows); by and large, the map of its antisymmetric component visually resembles that of the full $U_y$. This is reflected by the graphs of the two components of rms $U_y$: the characteristic rms values of the antisymmetric component are several times larger than the characteristic values of the symmetric component. These quantities do not exhibit dramatic changes in the course of the activity cycle; there is only a moderate depression in the amplitude of the antisymmetric component during the years of the activity maximum. The components of rms $(U_y)_\mathrm d$ do not differ so strongly. They vary similarly from the beginning of 2011 to the beginning of 2016, generally, differing by a factor of $\sim 1.5$--2 and following the variations in the sunspot number; however, as the beginning of Cycle 25 is approached, the symmetric component becomes relatively weak.

\section{Conclusion and Discussion}

We have seen that the zonal-flow field at depths of $d=0.5$--15.0~Mm exhibits alternating bands of increased and decreased velocities in the time--latitude diagrams. This pattern corresponds to torsional oscillations, or waves propagating toward the equator. One pair of high-speed bands stretches over the whole 10-year time interval, from latitudes of about $\pm 20\degree$ in 2010 to the equator at the beginning of 2020. The velocities in this pair reach a maximum in the southern band at the epoch of a sunspot-number peak, not far from the beginning of the year 2014. This velocity maximum agrees well in its latitudinal location with the maximum of magnetic fields. Another local maximum in the southern band is also distinguishable in the mid-2011, during a secondary activity peak. The second pair of increased-velocity bands converging to the equator can be traced starting from 2011--2012 in the southern hemisphere and, with a lesser certainty, from 2013--14 in the northern hemisphere. Zonal velocities are typically larger in the southern hemisphere. The dramatic growth of the zonal velocities in 2018 appears to be a precursor of the beginning of activity Cycle 25. The strong symmetrization by 2020 can be considered another precursor. At this stage, however, the amount of the available data is not sufficient to make such inferences with more certainty.

On the one hand, as we see, a pair of converging bands of `fast' rotation in the torsional-oscillation pattern extends over the whole interval considered, which is almost 10 year long; likely, the whole oscillation period is comparable in its length with a doubled solar-activity cycle and can be described as an extended solar cycle. The wedge-like features formed by the pairs of bands of `slow' and `fast'  rotation are nested in one another; a new wave emerges in the polar regions long before the disappearance of the preceding one near the equator. On the other hand, the zonal-velocity variations are definitely related to the solar-activity level, the local-velocity increases corresponding to the sunspot-number increases and being localized at latitudes where the strongest magnetic fields are recorded.

The general pattern of poleward meridional flow was superposed during the year of activity maximum with a higher harmonic corresponding to meridional flows directed toward the highest-activity latitudes. This flow component was more pronounced in the southern hemisphere; at levels $d=4.0$ and 8.5~Mm, it manifested itself in a reverse (equatorward) flow at relatively high latitudes that encountered a regularly directed flow at lower latitudes. In other words, a secondary flow converging to the spot-formation latitudes was present. During 2011 in the northern hemisphere and much longer in the southern hemisphere, the poleward-flow speed was increased at moderate depths and moderate latitudes, from $\varphi \sim 10\degree$ to $\varphi \sim 20\degree$. The $(U_y)_\mathrm d$ field becomes considerably closer to an antisymmetric one in 2018--2019, when the beginning of a new activity cycle approaches.

Our findings based on the time--distance helioseismic inversions reveal the extended-solar-cycle behavior of both the latitudinal migration of the zonal flows and the secondary meridional circulation component.
The flow variations originate at about 50--60\degree\ latitude during a solar maximum and migrate toward the equator during the following 22 years so that, at each moment, we observe a superposition of two wave-like migrating patterns. Our results show that the rms velocity amplitude of the antisymmetric component of the zonal flows remained relatively steady, varying between 0.5 and 1 m\,s$^{-1}$, while the symmetric component decreased during Cycle 24 from 2 to 1 m\,s$^{-1}$ but rapidly increased from 1 to 3~m\,s$^{-1}$ during the solar minimum in 2018--2020 (Figures~\ref{fig3}e,f). The local flow velocity reached 8~m\,s$^{-1}$. The meridional-flow velocity variations are mostly antisymmetric, corresponding to the general antisymmetric nature of the poleward circulation. The most significant deviations (`symmetric' component) are observed during the activity maximum (Figure~\ref{fig5}). During this period, the velocity of the mean meridional flow can substantially decrease and even be reversed. Such strong variations must be taken into account by the flux-transport theories.

In addition to the activity-relative variations, the meridional circulation exhibits longer-term variations similar to the extended-solar-cycle behavior of the zonal flows. This conclusion is consistent with the suggestion of \citet{Komm_etal_2020} that, in addition to the mechanism associated with inflows in active regions, there is ` a near-surface effect that depends on magnetic
activity being present below but not at the surface'. Our results show that the long-term variations, observed in the form of the additional circulation cells converging toward the active latitudes, sharply increased during the solar minimum of 2017--2020, prior to the start of Solar Cycle 25. A similar effect was observed by \citet{GonzalezHernandez_2010} prior to Solar Cycle 24. Therefore, this is an intrinsic feature of the solar dynamo.

Variations of the meridional circulation associated with the activity belts were predicted by the model of \citet{Spruit_2003}, who suggested that the zonal- and meridional-flow variations are driven by temperature variations near the surface due to the enhanced emission of radiation by the small-scale magnetic-field elements in active regions. This model is faced with the difficulty that the migrating flow patterns are observed and become stronger during the activity minima.

Studies by \citet{Lekshmi_etal_2018,Lekshmi_etal_2019} were specifically focused on the hemispherical asymmetry of the zonal and meridional flows determined helioseismologically. As a measure of the asymmetry of a quantity $U$, the authors used the difference $U_\mathrm{North}U_\mathrm{South}$ between the $U$ values in the two hemispheres at a given latitude and a given depth. The zonal flows were found to exhibit a positive correlation between the asymmetry of $(U_x)_\mathrm{d}$ and the asymmetry of solar activity measured by the sunspot number and total magnetic flux, the $(U_x)_\mathrm{d}$ asymmetry preceding the activity by 1--1.5 yr or so
\citep{Lekshmi_etal_2018}. As for the meridional flow, the asymmetry in $(U_y)_\mathrm{d}$ exhibits an anticorrelation with the solar-activity asymmetry, being ahead of the latter by 3.1--3.5 yr \citep{Lekshmi_etal_2019}.

We considered here the variations in the velocity asymmetry with the level, rather than asymmetry, of solar activity. The differential rotation is normally symmetric about the equator (i.e., the zonal velocity is an even function of latitude). To make comparisons with the results obtained by Lekshmi et al.,
we should observe that the differential rotation became by 2020 substantially more symmetric (with a symmetric-to-antisymmetric $(U_x)_\mathrm{d}$-amplitude ratio of about 6) than it was in 2017 (with such a ratio close to 1). The hemispherical asymmetry of solar activity seems to have more subtle effects on
the differential-rotation asymmetry.

In contrast to the zonal flow, the meridional-flow velocity is typically antisymmetric with respect to the equator (reflecting the flow direction from the equator to the poles). Accordingly, its closeness to this normal flow pattern can be characterized by the antisymmetric-to-symmetric $(U_y)_\mathrm{d}$-amplitude ratio. During 2010--2017, this ratio mainly varied between 1.5 and 2, following the variation in the sunspot number. Since 2016, it grew and reached values of 4-6 by the beginning of the new activity cycle.

\citet{Zhao_Nagashima_etal_2012,Gizon_2020}  
 and others noted systematic center-to-limb variations in the measured helioseismic travel times, which can introduce errors in the determination results for interior meridional flows. However, these variations mostly affect the determination of the flow structure in the deep convection zone, and are not significant for the measurements in the relatively shallow subsurface layers, presented here. To evaluate this effect, we additionally calculated the meridional velocities using data for Stonyhurst-longitude ranges of $\pm 40\degree$ and $\pm 20\degree$. The results do not exhibit any appreciable dependence on the range used. The diagrams of Figures 4 and 5 are visually indistinguishable between the cases of different ranges, the rms $(U_y)_\mathrm{d}$ differences not exceeding 0.1 m\,s$^{-1}$ in order of magnitude.

A recent non-linear dynamo model of \citet{Pipin_Kosovichev_2019} explains the observed `extended-cycle' flow variations with modulation of convective heat flux by the dynamo-generated magnetic field deep in the convection zone. The model predicts the extended solar-cycle behavior of both the zonal and meridional flows. Thus, it is supported by the results presented in this paper, as well as by the global-helioseismology analysis of the spatiotemporal evolution of the deep zonal flows, which revealed a pattern of migrating dynamo waves \citep{Kosovichev_Pipin_2019}. Further model calculations \citep{Pipin_2020} found a rather weak correlation between the surface-zonal-flow amplitude and strength of the following activity cycle. Therefore, it is still unclear how the recent sharp increase in the flow speed is related to the strength of the upcoming solar maximum.

\acknowledgments

The helioseismological data used here were derived from HMI observational data available courtesy of the NASA/SDO and HMI science teams. We also used sunspot-number data from the World Data Center for Sunspot Index, and Long-term Solar Observations (WDC-SILSO), Royal Observatory of Belgium, Brussels. The work partially supported by NASA grants: NNX14AB70G, 80NSSC20K1320, 80NSSC20K0602.

\bibliography{Getling}
\end{document}